\documentclass[aps,pra,superscriptaddress,showpacs,twocolumn]{revtex4-2}
\usepackage[left=1.9cm,right=1.9cm,bottom=1.9cm,top=1.9cm,ignoreall]{geometry}
\usepackage[utf8]{inputenc}
\usepackage{graphicx}
\usepackage{amsmath}
\usepackage{amssymb}
\usepackage{xspace}
\usepackage{bm}
\usepackage{color}
\usepackage[squaren,Gray]{SIunits}
\usepackage[hyperindex=true]{hyperref}
\hypersetup{linktocpage,colorlinks=true,citecolor=blue,linkcolor=blue}
\usepackage{empheq}
\usepackage{braket}
\usepackage{physics}

\begin{document}
	
\title{On-chip generation of hybrid polarization-frequency entangled biphoton states}

\author{S. Francesconi}
\affiliation{Université Paris Cité, CNRS, Laboratoire Matériaux et Phénomènes Quantiques, 75013 Paris, France}

\author{A. Raymond}
\affiliation{Université Paris Cité, CNRS, Laboratoire Matériaux et Phénomènes Quantiques, 75013 Paris, France}

\author{R. Duhamel}
\affiliation{Université Paris Cité, CNRS, Laboratoire Matériaux et Phénomènes Quantiques, 75013 Paris, France}

\author{P. Filloux}
\affiliation{Université Paris Cité, CNRS, Laboratoire Matériaux et Phénomènes Quantiques, 75013 Paris, France}

\author{A.~Lemaître}
\affiliation{Université Paris-Saclay, CNRS, Centre de Nanosciences et de Nanotechnologies, 91120, Palaiseau, France}

\author{P. Milman}
\affiliation{Université Paris Cité, CNRS, Laboratoire Matériaux et Phénomènes Quantiques, 75013 Paris, France}

\author{M. I. Amanti}
\affiliation{Université Paris Cité, CNRS, Laboratoire Matériaux et Phénomènes Quantiques, 75013 Paris, France}

\author{F. Baboux}
\thanks{Corresponding author: florent.baboux@u-paris.fr}
\affiliation{Université Paris Cité, CNRS, Laboratoire Matériaux et Phénomènes Quantiques, 75013 Paris, France}

\author{S. Ducci}
\affiliation{Université Paris Cité, CNRS, Laboratoire Matériaux et Phénomènes Quantiques, 75013 Paris, France}

\date{\today}

\begin{abstract}

We demonstrate a chip-integrated semiconductor source that combines polarization and frequency entanglement, allowing the generation of entangled biphoton states in a hybrid degree of freedom without postmanipulation.
Our AlGaAs device is based on type-II spontaneous parametric down-conversion (SPDC) in a counterpropagating phase-matching scheme, in which the modal birefringence lifts the degeneracy between the two possible nonlinear interactions. This allows the direct generation of polarization-frequency entangled photons, at room temperature and telecom wavelength, and in two distinct spatial modes, offering enhanced flexibility for quantum information protocols. The state entanglement is quantified by a combined measurement of the joint spectrum and Hong-ou-Mandel interference of the biphotons, allowing to reconstruct a restricted density matrix in the hybrid polarization-frequency space.

\end{abstract}

\maketitle

\section{Introduction}

Quantum states of light are central resources for quantum information technologies. Indeed, besides their easy transmission and robustness to decoherence, photons provide a large variety of degrees of freedom (DOF) to encode information, which can be either two-dimensional (such as polarization) or higher-dimensional (such as frequency, orbital angular momentum or spatial modes)~\cite{Walmsley15,Flamini19}. In addition, photonic information can either be encoded in individual photons or in the quadratures of the electromagnetic field, defining respectively the realms of discrete variable (DV) and continuous variable (CV) encoding.

Polarization is a paradigmatic two-dimensional photonic DOF that allowed for pioneering demonstrations in quantum information, ranging from fundamental tests of quantum mechanics \cite{Aspect81} to quantum computing \cite{Shor97} and communication tasks \cite{Ekert91,Bouwmeester97}.
Focusing on DV encoding, polarization Bell states, such as $\ket{\Psi^+}_{\rm polar}=\left (\ket{H}\ket{V}+\ket{V}\ket{H} \right) / \sqrt{2} $, where $H$ and $V$ stand for the horizontal and vertical polarizations of single photons, constitute a fundamental building block for many of these applications. They can be efficiently generated with parametric processes in nonlinear bulk crystals combined with external components (such as a walk-off compensator or a Sagnac interferometer)  \cite{Kwiat95,Kwiat99,Shi04,Kim06}. More recently, chip-based sources based on quantum dots \cite{Dousse10,Muller14,Huber17,Jons17,Liu19b} or parametric processes \cite{Matsuda12,Valles13,Horn13,Orieux13,Sansoni17} allowed to generate polarization Bell states in a fully integrated manner, without requiring external optical elements.

\begin{figure*}[t]
\centering
\includegraphics[width=0.9\textwidth]{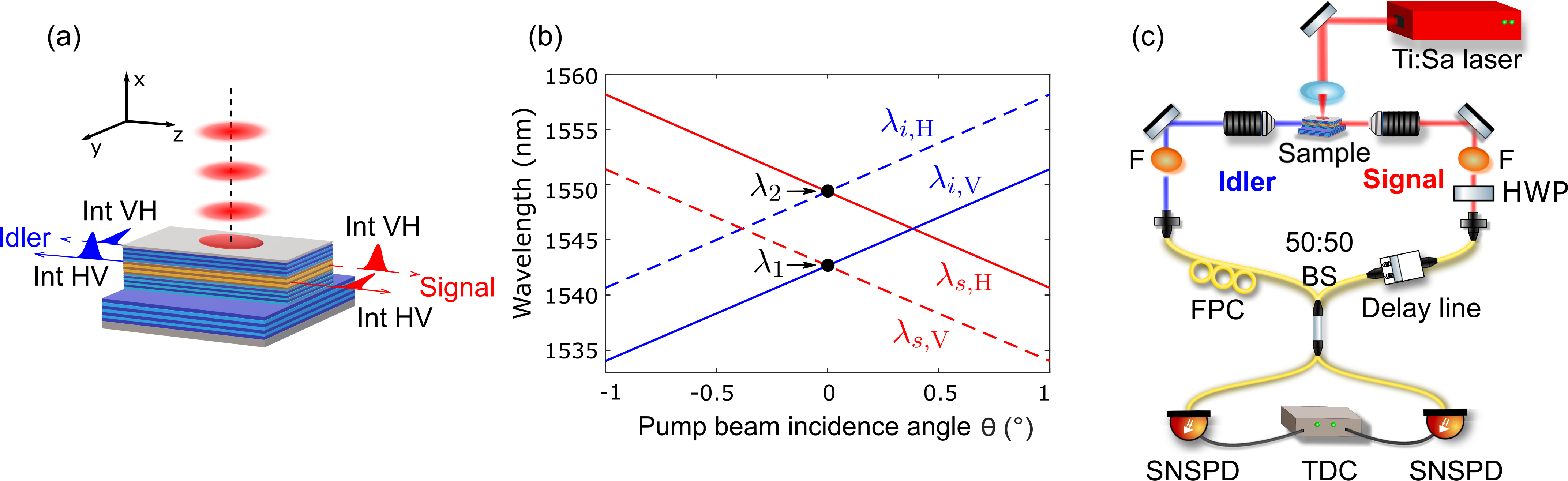}
\caption{
(a) Schematics of an AlGaAs ridge microcavity emitting counterpropagating twin photons by SPDC in a transverse pump geometry. Two type-II interactions occur, generating either an H polarized signal photon and a V polarized idler photon (interaction $HV$), or the opposite situation (interaction $V\!H$), resulting in a hybrid polarization-frequency entangled state.
(b) Calculated SPDC tunability curve, showing the central wavelengths of signal and idler photons as a function of the pump incidence angle $\theta$ (with respect to the vertical $x$ axis), for both interactions, using our sample properties and pump wavelength $\lambda_p=773.15$ mn.
(c) Sketch of the experimental setup to measure the HOM interference of the hybrid polarization-frequency state (abbreviations: HWP = half-wave plate, F = frequency filter, SNSPD = superconducting nanowire single-photon detector, TDC = time-to-digital converter).
}
\label{Fig1}
\end{figure*}

Now turning to high-dimensional photonic DOF, among the various candidates, frequency is attracting a growing interest due to its robustness to propagation in optical fibers and its capability to convey large scale quantum information into a single spatial mode.
Frequency is intrinsically a continuous DOF, that can be used to encode information as such \cite{Donohue16,Ansari18b,MacLean18}, but it can also be viewed as a discrete DOF when divided into frequency bins \cite{Olislager10,Kues19}.
In the latter case, the simplest maximally entangled state of two photons is the so-called two-colour Bell state, $\ket{\Psi^+}_{\rm col}=\left (\ket{\omega_1}\ket{\omega_2}+\ket{\omega_2}\ket{\omega_1} \right) / \sqrt{2} $,
where $\ket{\omega_{1}}$ and $\ket{\omega_{2}}$ are well separated single-photon frequency bins. 
Several experimental schemes have been implemented to generate such two-color entangled states, which could be exploited e.g. as a metrology resource for precise time measurements \cite{Chen19b}, or to interconnect stationary qubits with dissimilar energy levels \cite{Olmschenk09} in a quantum network.
The first realizations relied on filtering out frequency bins from a continuous spectrum \cite{Ou88, Rarity90}. More recently, brighter sources have been demonstrated by using periodically poled crystals in crossed configurations \cite{Kaneda19, Chen19b}, Sagnac loops \cite{Li09}, double passage configurations \cite{Jin18}, or by transferring entanglement from the polarization to the frequency domain \cite{Ramelow09}. All these demonstrations relied on bulk nonlinear crystals, and while integrated sources such as microring resonators are powerful to generate frequency combs (involving a high number of frequency bins) \cite{Kues19}, the direct and versatile generation of two-color entangled states with chip-based sources is still scarce. In the latter domain an important advance was achieved in Ref. \cite{Silverstone14}, combining on the same Silicon chip 2 four-wave mixing sources and an interferometer with reconfigurable phase shifter. The resulting interference between two independent sources allowed generating two-color entangled states in an integrated and controlled manner, albeit with a limited efficiency.

Besides entanglement into a single degree of freedom, combining several DOF can provide an increased flexibility for quantum information protocols.
To this aim, we demonstrate here a single chip-integrated semiconductor source that combines frequency and polarization entanglement, leading to the generation of hybrid polarization-frequency entangled biphoton states without post-manipulation.
Our AlGaAs device is based on type-II spontaneous parametric down-conversion (SPDC) in a counterpropagating phase matching scheme, where the modal birefringence lifts the spectral degeneracy between the two possible nonlinear interactions occurring in the device. This allows the direct generation of polarization-frequency entangled photons in two distinct spatial modes, at room temperature and telecom wavelength. 
Such combination of degrees of freedom opens enlarged capabilities for quantum information applications, allowing to switch from one DOF to another and thus to adapt to different experimental conditions in a versatile manner.

\section{Theoretical framework}

Our semiconductor integrated source of photons pairs is sketched in Fig. \ref{Fig1}a. It is a Bragg ridge microcavity made of a stacking of AlGaAs layers with alternating aluminum concentrations \cite{Caillet09,Orieux11,Orieux13}. The source is based on a transverse pump scheme, where a pulsed laser beam impinging on top of the waveguide (with incidence angle $\theta$ with respect to the $x$ axis) generates pairs of counterpropagating photons (signal and idler) through SPDC \cite{DeRossi02,Orieux13}. 
As a consequence of the opposite propagation directions for the photons, two type-II SPDC processes occur simultaneously in the device \cite{Orieux13}: a first one that generates a TE-polarized signal photon (propagating along $z>0$, see Fig.\ref{Fig1}a) and a TM-polarized idler photon (propagating along $z<0$); and a second one that generates a TM-polarized signal and a TE-polarized idler. We later refer to these two generation processes as $HV$ and $V\!H$ respectively, using the shorter notation $H$ (horizontal) for TE and $V$ (vertical) for TM. 
The central frequencies $\omega_s$ and $\omega_i$ of the signal and idler photons obey energy conservation ($\omega_p=\omega_s+\omega_i$, where $\omega_p$ is the pump frequency) and momentum conservation along the waveguide direction, which reads for the two interactions: 
\begin{align}\label{EqPM}
	&\omega_p \sin(\theta)=\omega_s n_H (\omega_s) - \omega_i n_V (\omega_i)  &\text{\quad [Inter. HV]} \\
	&\omega_p \sin(\theta)=\omega_i n_V (\omega_s) - \omega_i n_H (\omega_i)  &\text{\quad [Inter. VH]} 
\end{align}
with $n_H$ and $n_V$ the modal refractive indices of the waveguide for the $H$ and $V$ polarizations respectively.

Figure \ref{Fig1}b shows the resulting SPDC tunability curve, i.e. the  calculated central wavelengths of signal and idler photons as a function of the pump incidence angle $\theta$, for both interactions, by taking into account our sample properties and the used pump wavelength $\lambda_p=773.15$ mn. 
The two interactions are not degenerate because of the small modal birefringence of the waveguide ($\Delta n=n_H-n_V \simeq 1.2\times 10^{-2}$ at the working temperature $293$ K).
In the low-pumping regime, the generated two-photon state resulting from both interactions reads 
\begin{equation}
\begin{aligned}
	\ket{\Psi}=\iint \dd\omega \dd\omega' \big[ 
	&\phi_{HV} (\omega, \omega') \hat{a}^{\dagger}_{s,H}(\omega)\hat{a}^{\dagger}_{i,V}(\omega') \\
	+ &\phi_{VH} (\omega, \omega') \hat{a}^{\dagger}_{s,V}(\omega)\hat{a}^{\dagger}_{i,H}(\omega') \big] \ket{0,0}
\label{Eq:Psi}
\end{aligned}
\end{equation}
where the operator $\hat{a}^\dagger_{s(i),H(V)}(\omega)$ creates a signal (idler) photon of frequency $\omega$ and polarization $H$ ($V$). The function $\phi_{HV} (\omega, \omega')$ is the Joint Spectral Amplitude (JSA) for the interaction \textit{HV}, i.e. the probability amplitude to measure an $H$ signal photon at frequency $\omega$ and a $V$ idler photon at frequency $\omega'$; the analogous definition goes for $\phi_{VH}$.

\begin{figure}[h]
\centering
\includegraphics[width=1\columnwidth]{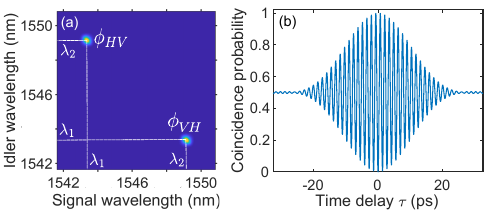}
\caption{
(a) Simulated joint spectral intensity (JSI) of the hybrid polarization-frequency biphoton state of Eq. \eqref{Eq:Psi}, assuming Gaussian phase-matching functions (see text for details).
(b) Simulated HOM interferogram, showing a sinusoidal oscillation modulated by a Gaussian envelope.}
\label{Fig2}
\end{figure}

To produce polarization-frequency entangled states, we consider the situation where the pump beam impinges at normal incidence on the waveguide ($\theta=0$).
The two interactions give rise to two distinct peaks in the biphoton spectrum, as seen in the simulation of Fig. \ref{Fig2}a showing the Joint Spectral Intensity (JSI), which is the modulus squared of the JSA (plotted in the wavelength space).
The two peaks, centered at wavelengths $\lambda_1$ and $\lambda_2$ (see Fig. \ref{Fig1}b), are symmetric with respect to the degeneracy wavelength $\lambda_{\rm deg}=2 \lambda_p$.
Since the separation between the peaks is much larger than their spectral width, a reasonable approximation is to discretize the frequency degree of freedom and replace the JSAs by Dirac deltas,  $\phi_{HV}(\omega, \omega')=\phi_{VH}(\omega', \omega)=\delta(\omega-\omega_1, \omega'-\omega_2)$ (with $\omega_{1(2)}=2\pi c / \lambda_{1(2)}$). Inserting these expressions in Eq.~\eqref{Eq:Psi}, the emitted biphoton state can be rewritten as a hybrid polarization-frequency (HPF) state, $\ket{\Psi} \simeq \ket{\Psi}_{\rm HPF}$, with
\begin{equation}
\ket{\Psi}_{\rm HPF} = \frac{1}{\sqrt{2}} \left (\ket{H \omega_1}_s\ket{V \omega_2}_i+\ket{V \omega_2}_s\ket{H \omega_1}_i \right)
\label{Eq:Psi_discrete}
\end{equation}
The first ket represents the signal photon and the second the idler photon, as defined by their respective opposite propagation directions.
In this state, the frequency $\omega_1$ is always associated to the $H$ polarization, while the frequency $\omega_2$ is always associated to the $V$ polarization. This constitutes a composite polarization-frequency degree of freedom, and the state $\eqref{Eq:Psi_discrete}$ is maximally entangled in this composite degree of freedom. 
An advantage of such kind of entangled state is its versatility, since upon manipulation with simple optical elements, the state $\eqref{Eq:Psi_discrete}$ can be projected either on a polarization Bell state \cite{Kim03} or a two-color Bell state \cite{Ramelow09}, so as to adapt to a specific experimental context.

To reveal and quantify the entanglement level of the HPF state, two-photon interference in a Hong-Ou-Mandel (HOM) experiment provides a powerful tool, as it directly probes the quantum coherence between the two components of the state.
When the signal and idler photons are delayed by a time $\tau$ and sent in the two input ports of a balanced beamsplitter, the coincidence probability between the beamsplitter outputs can be calculated from the JSAs $\phi_{HV}$ and $\phi_{VH}$ of the two SPDC processes: 
\begin{equation}
	\resizebox{\hsize}{!}{$P_c(\tau)=\dfrac{1}{2}-\Re  \qty[\iint \dd\omega \dd\omega' \phi_{HV} (\omega, \omega') \phi_{VH}^* (\omega', \omega)  e^{-i (\omega-\omega' )\tau}]$}
	\label{Eq:HOM_1}
\end{equation}
where "Re" denotes the real part. Coming back to the continuous (rather than Dirac delta) expression of the joint spectra, and considering a narrow pump bandwidth and negligible group velocity dispersion (as justified in our experimental conditions), the JSAs can be written as $\phi_{\alpha} (\omega,\omega')=\phi_{\alpha}^{\rm spec}(\omega_+) \, \phi_{\alpha}^{\rm pm}(\omega_-)$ (with $\alpha=$ $HV$ or $VH$), where we have introduced $\omega_\pm=\omega \pm \omega'$ \cite{Boucher15,Barbieri17}.
The function $\phi_{\alpha}^{\rm spec}$, corresponding to the condition of energy conservation, is given by the spectrum of the pump beam, while the function $\phi_{\alpha}^{\rm pm}$, reflecting the phase-matching condition, is determined by the spatial properties of the pump beam \cite{Orieux13,Boucher15}. Due to the separability of the JSAs in $\omega_+$ and $\omega_-$ the HOM coincidence probability is actually determined only by the phase-matching part component of the JSAs \cite{Douce13,Barbieri17}. Considering a Gaussian pump spot of waist $w_z$ along the waveguide direction, the phase-matching functions of the two interactions can be calculated as:
\begin{equation}
	\begin{split}
		& \phi_{HV}^{\rm pm}(\omega_-)= \sqrt\pi \: w_z \: e^{-(\omega_- - \mu)^2 / 2 \Delta \omega_-^2} \\
		& \phi_{VH}^{\rm pm}(\omega_-)= \sqrt\pi \: w_z \: e^{-(\omega_- + \mu)^2 / 2 \Delta \omega_-^2}
	\end{split}
\label{Eq:PM}
\end{equation}
where $\Delta \omega_-=\sqrt{2} \,v_g / w_z$ is the spectral width of each interaction and $\mu = \omega_1 - \omega_2 = v_g \, \omega_p (n_H-n_V)  / 2c$ their spectral separation, with $v_g$ the harmonic mean of the group velocities of the SPDC modes \cite{Boucher15}. Inserting Eqs. \eqref{Eq:PM} into the expression of the coincidence probability \eqref{Eq:HOM_1} leads to:
\begin{equation}
	P_c(\tau)=\dfrac{1}{2}-\dfrac{1}{2} \exp[-\dfrac{\tau^2}{2 \Delta\tau^2}] \cos(\mu \tau)	
	\label{Eq:HOM_2}	
\end{equation}
The resulting interferogram, plotted in Fig. \ref{Fig2}b, displays a sinusoidal oscillation (spatial quantum beating), of frequency $\mu$ equal to the spectral separation of the frequency modes, and modulated by a Gaussian envelope of width $\Delta\tau =\sqrt{2} /\Delta \omega_-$ determined by the spectral width of each interaction.

\section{Experiments}

The epitaxial structure of the sample is made of a 4.5-period Al$_{0.80}$Ga$_{0.20}$As/Al$_{0.25}$Ga$_{0.75}$As core, surrounded by two distributed Bragg mirrors made of a 36- and 14-period Al$_{0.90}$Ga$_{0.10}$As/Al$_{0.35}$Ga$_{0.65}$As stacking for the bottom and top mirrors, respectively. The Bragg mirrors provide both a vertical microcavity to enhance the pump field and a cladding
for the twin-photon modes. From this planar structure a ridge waveguide (length $2.6$ mm, width 5 $\mu$m and height 7 $\mu$m) is fabricated by electron beam lithography followed by wet etching. The waveguide facets are then coated with a thin SiO$_2$ film (target thickness $270$ nm) deposited by PECVD, resulting in a $\simeq 10 \%$ modal reflectivity for the SPDC modes.

The experimental setup is shown in Fig. \ref{Fig1}c. The sample is pumped with a pulsed Ti:Sa laser of central wavelength $\lambda_p \simeq 773.15 $ nm, pulse duration $4.5$ ps, repetition rate $76$ MHz and average power $30$ mW on the sample. A cylindrical lens focuses the pump beam into a Gaussian elliptical spot on the top of the waveguide (waist $w_z=1$ mm along the waveguide direction) at perpendicular incidence ($\theta=0$), and the generated infrared photons are collected with two microscope objectives and collimated into single-mode optical fibers.

\begin{figure*}[t]
	\centering
	\includegraphics[width=0.8\textwidth]{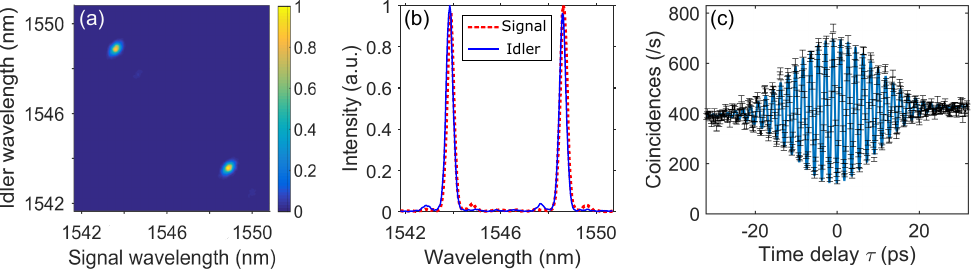}
	\caption{
		(a) Measured joint spectral intensity (JSI) of the hybrid polarization-frequency state, and (b) corresponding marginal spectrum of signal (red) and idler (blue) photons.
		(c) Measured HOM interferogram (black symbols with error bars) fitted with Eq. \eqref{Eq:HOM_3} (blue line). Data points show raw (uncorrected) coincidence counts.
	}
	\label{Fig3}
\end{figure*}

We first characterize the generated quantum state by measuring the JSI using a fiber spectrograph \cite{Eckstein14}. For this, the signal and idler photons are separately sent into a spool of highly dispersive fiber, converting the frequency information into a time-of-arrival information. The latter is recorded using superconducting nanowire single photon detectors (SNSPD, of detection efficiency $90\%$) connected to a time-to-digital converter (TDC); long-pass filters are used to remove slight luminescence noise from the sample.
The measured JSI, reported in Fig. \ref{Fig3}a, shows two well-defined frequency peaks, symmetric with respect to the degeneracy wavelength, in good qualitative agreement with the numerical simulation of Fig. \ref{Fig2}a. This can be better seen in Fig. \ref{Fig3}b, showing the marginal spectrum of signal (red) and idler (blue) photons as extracted from the experimental JSI. The frequency peaks have a FWHM of $\simeq 0.5$ nm and separation $\Delta \lambda=\lambda_2-\lambda_1 \simeq 4.8$ nm, i.e. about $10$ times higher than their linewidth. The measured $\Delta \lambda$ is smaller than in the simulation ($6.2$ nm in Fig. \ref{Fig2}a), pointing to a discrepancy between the experimental and simulated modal birefringence $\Delta n=n_H-n_V $. This could be due to a slight deviation of the epitaxial structure from the nominal one and/or imperfections of the simulation (used material refractive indices and exact etching shape of the waveguide).

We now perform two-photon interference in a HOM setup (see Fig. \ref{Fig1}c) to reveal the entanglement properties of the generated hybrid polarization-frequency state. A half-wave plate (HWP) in the signal arm and a fibered polarization controller (FPC) in the idler arm are used to compensate for polarization rotation on the optical path, hence maintaining the signal and idler photons in the same state than at the chip output so that they enter the beamsplitter with crossed polarizations.
The resulting interferogram, shown in Fig. \ref{Fig3}c (black points with error bars), displays a clear sinusoidal modulation with a Gaussian envelope. Each point is obtained by a 20 s integration time and error bars are calculated assuming a Poissonian statistics. The data is fitted (blue line) using a modified version of Eq. \eqref{Eq:HOM_2} accounting for experimental imperfections:
\begin{equation}
P_c(\tau)=\dfrac{1}{2}\left (1 - V \exp[-\dfrac{\tau^2}{2 \Delta\tau^2}] \cos(\mu \tau)\right ) +a \tau +b
\label{Eq:HOM_3}	
\end{equation}
where $V$ is the interferogram visibility, and the linear term $a \tau +b$ accounts for a slight drift of the alignment during the total time of the measurement.
The fitted envelope width is $\Delta\tau=10.0$ ps, in good agreement with the simulation ($10.5$ ps) of Fig. \ref{Fig2}b. The fitted oscillation period is $2 \pi / \mu =1.8$ ps, in good agreement with the spectral measurement of the peaks separation $\mu$ (Fig. \ref{Fig3}b), but higher than in the simulation ($1.3$ ps), for the same reasons than mentioned above.

The experimental raw visibility is $V=70.1 \pm 1.1 \%$. We attribute the main part of the visibility reduction to the non-zero reflectivity of the facets. Indeed, as a consequence of the latter, the two photons of each pair have a non-zero probability to exit through the same facet, instead of opposite ones. This results in a Franson-type interference \cite{Franson89} between the situation where both photons exit from one facet and the situation where both photons exist from the other facet, leading to a modulation at a frequency equal to the sum of signal and idler frequencies, i.e. the pump frequency $\omega_p$. Here this Franson-type interference occurs for a time delay shorter than the photon coherence time and therefore it is superimposed to the HOM interference \cite{Halder05}. The corresponding period is $2\pi / \omega_p \simeq 2 $ fs, which is beyond the resolution of our HOM setup. The measured interferogram thus averages out over these rapid Franson oscillations, reducing the effective visibility of the fringes. For our sample with $10 \%$ reflectivity for the SPDC modes, numerical simulations including this averaging effect predict a visibility of $82 \%$ (see Appendix for details). We attribute the remaining visibility drop to slight imperfections of the pump spatial profile and incidence angle \cite{Francesconi20}.

\begin{figure}[t]
\centering
\includegraphics[width=0.8\columnwidth]{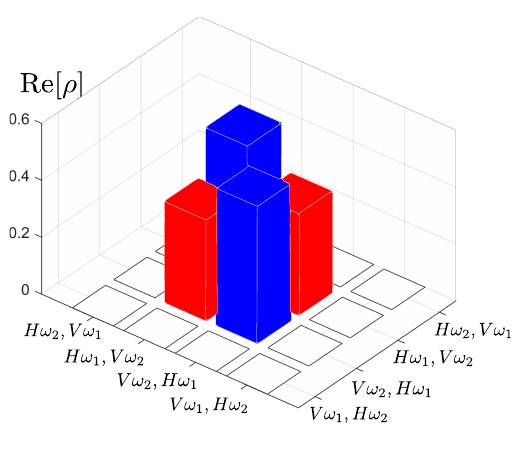}
\caption{
Experimental reconstruction of the restricted density matrix (Eq. \eqref{eq:matrix}) of the biphoton state in the hybrid polarization-frequency discrete space (the imaginary part is zero by construction).
}
\label{Fig4}
\end{figure}

Using the joint spectrum and HOM measurements we can now quantify the entanglement of the generated HPF state by estimating a restricted density matrix \cite{Ramelow09} in the hybrid polarization-frequency discrete space. The full basis would include all combinations of the \{$H,V$\} polarizations and \{$\omega_1,\omega_2$\} frequencies, resulting in a 16x16 density matrix. However, physical considerations allow restricting the relevant basis dimension, projecting it onto the relevant subspace.
Indeed, the employed type-II SPDC process does not allow the production of photons of same polarization, while energy conservation forbids the production of photons of same frequency; in addition, the phase-matching imposes that photons of frequency $\omega_1$ (resp. $\omega_2$) are always $V$ (resp. $H$) polarized. This leads to the following 4x4 restricted density matrix:
\begin{equation}
\rho=\mqty(0 & 0 & 0 & 0 \\ 0 & p & \dfrac{V}{2} e^{i \varphi} & 0 \\ 0 & \dfrac{V}{2} e^{-i \varphi} & 1-p & 0 \\ 0 & 0 & 0 & 0) 
\label{eq:matrix}
\end{equation}
expressed in the basis $\{\ket{H\omega_2}_{\!s} \!\ket{V \omega_1}_{\!i}$, $\ket{H\omega_1}_{\!s} \! \ket{V \omega_2}_{\!i}$, $\ket{V\omega_2}_{\!s} \! \ket{H \omega_1}_{\!i}$, $\ket{V\omega_1}_{\!s} \ket{H \omega_2}_{\!i}\}$.
The parameters $p$ (balance parameter) and $V$ (visibility) are real and obey the physical constraints $0\leq p \leq 1$ and ${0\leq V/2 \leq \sqrt{p(1-p)}}$ \cite{Ramelow09}.

The JSI measurement (Fig. \ref{Fig3}a) yields the population term $p=0.517 \pm 0.005 $ while the HOM interferogram (Fig. \ref{Fig3}c) gives the coherence modulus $V/2$ from the visibility deduced above; the phase $\varphi=0$ between the two interactions is deduced from the fact that the source is pumped by a single pump beam. 
The resulting density matrix is shown in Fig. \ref{Fig4}. 
It allows to extract the purity of the generated state, $P=0.746 \pm 0.008$, 
its fidelity to the ideal state of Eq. \eqref{Eq:Psi_discrete}, $F=0.851 \pm 0.007$,
and its concurrence, $C=0.701 \pm 0.011$ (all raw values). This confirms the direct generation of hybrid polarization-frequency entanglement by our chip-integrated source. The generation rate, estimated from single and coincidence counts data \cite{Tanzilli02}, is $\simeq 10^7$ pairs/s at the chip output.

\section{Discussion and conclusion}

In summary, we have demonstrated a chip-based semiconductor source that combines polarization and frequency entanglement, enabling the generation of hybrid polarization-frequency entangled photon pairs directly at the generation stage, in two distinct spatial modes.
Such combination of degrees of freedom provides an increased flexibility for quantum information protocols, allowing to adapt the source to different applications in a versatile manner. The demonstrated device operates at room temperature and telecom wavelength, is compliant with electrical pumping \cite{Boitier14} thanks to the direct bandgap of AlGaAs, and has a strong potential for integration within photonic circuits \cite{Wang14,Belhassen18,Dietrich16}.

These results could be further expanded along several directions. First, the fidelity of the experimentally generated state to the ideal HPF state of Eq. \eqref{Eq:Psi_discrete} could be improved by implementing a multi-layer coating of the waveguide facets, allowing to reach modal reflectivities $\leqslant 1 \%$. This enhancement would lead to an expected fidelity larger than $0.93$, all other factors (including experimental imperfections) kept unchanged. The fidelity could be further improved by correcting for the small imperfections of the pump spatial profile (deviations from the ideal Gaussian shape) using e.g. a spatial light modulator.

In addition, the frequency entanglement of our hybrid polarization-frequency state can be varied by different means. In Eq. \eqref{Eq:Psi_discrete}, frequency entanglement is described as a discrete two-color entanglement, which reflects the dominant frequency anticorrelation of the state, but neglects intra-mode frequency entanglement, i.e. the continuous entanglement associated to the internal structure of each frequency mode (as determined by the JSAs $\phi_{HV}$ and $\phi_{VH}$ in Eq. \eqref{Eq:Psi}). This intra-mode entanglement, which manifests in the envelope of the HOM oscillations (Eq. \eqref{Eq:HOM_2}), can be controlled \textit{in situ}. Indeed, a specific asset of our counterpropagating phase-matched source is that the JSA can be engineered through the spatial properties of the pump beam \cite{Boucher15}. We have here implemented Gaussian JSAs (Eq. \eqref{Eq:PM}), which leads to a Gaussian envelope in the HOM interferogram; the width of this envelope could be varied by changing the waist of the pump beam. Going beyond, more complex types of intra-mode entanglement could be designed by tailoring the spatial phase of the pump beam \cite{Francesconi20,Francesconi21}, leading to various envelope shapes which could be exploited e.g. for quantum metrology based on HOM interferometry \cite{Lyons18,Chen19b}.

Frequency entanglement can also be tailored at the inter-mode level, i.e. by varying the separation between the two central frequencies ($\omega_{1}$ and $\omega_{2}$). This can be achieved by playing with the modal birefringence of the waveguide, either \textit{in situ} by changing the working temperature, or at the fabrication step by varying the width of the waveguide. Numerical simulations show that the birefringence  $\Delta n=n_H-n_V$ decreases as the waveguide width is decreased, reaching zero for a width of $\simeq 1.3 \, \mu$m. In the latter situation, the two interactions become spectrally degenerate and the two frequency modes collapse. Thus, inter-mode frequency entanglement vanishes, but intra-mode entanglement, as determined by the JSA, remains and is decoupled from polarization entanglement. This would lead to an hyper-entangled polarization-frequency state, in the tensor product form
$\left (\ket{H}_s\ket{V}_i+\ket{V}_s\ket{H}_i \right) / \sqrt{2} \otimes \iint \dd\omega \dd\omega' \phi (\omega, \omega') \ket{\omega}_s \ket{\omega'}_i$. 
This situation opens stimulating perspectives for the implementation of quantum information tasks \cite{Kwiat97,Xie15,Deng17}, in particular in the field of quantum communication to improve bit rates and resilience to noise \cite{Steinlechner17,Vergyris19,Ecker19,Kim21}.

\section{Appendix: HOM interferogram with Fabry-Pérot effect}

We state in the article that the non-zero reflectivity of the waveguide facets is the main source of limitation of visibility in the measured HOM interferogram of the HPF state (Fig. \ref{Fig3}). We here demonstrate it by expanding our theoretical treatment to account for the Fabry-Pérot cavity effect induced by the facets.

\begin{figure}[h]
	\centering
	\includegraphics[width=0.64\columnwidth]{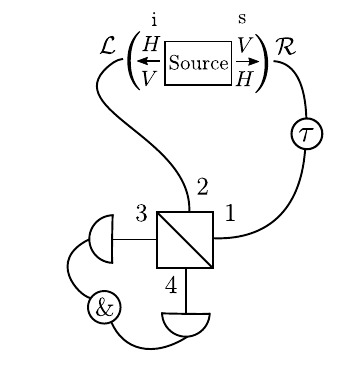}
	\caption{Hong-Ou-Mandel scheme for a counter-propagating parametric source emitting photons through both $HV$ and $VH$ interactions and considering the cavity effect induced by the waveguide facets. The letters refer to the subscripts used in the calculations.}
	\label{HOM_SchemeFP}
\end{figure}

We model the HOM experiment as shown in Fig. \ref{HOM_SchemeFP}, where the letters define the subscripts used in the calculations that follow. The source generates signal and idler photons: $s$ and $i$ label the propagation directions of the generated photons inside the source before the mixing by the Fabry-Pérot cavity (materialized by mirrors in the figure), while $\mathcal{R}$ (right) and $\mathcal{L}$ (left) denote the propagation directions outside the cavity. The ports $1$ and $2$ are the two inputs of the beamsplitter, $3$ and $4$ are the two outputs between which coincidences are measured. A delay line is placed is the $\mathcal{R}$ path.

We start from the expression of the emitted state without cavity effect (Eq. \eqref{Eq:Psi}). The reflection and transmission coefficients (in amplitude) of the cavity are written respectively as:
\begin{equation}
	\begin{split}
		&  f_r(\omega)= \dfrac{\sqrt{R(1-R)} \: \exp(i\dfrac{3\omega n L}{2 c})}{1-R \:\exp(i \dfrac{2\omega n L}{c} )} \\
		& f_t(\omega)= \dfrac{\sqrt{1-R} \: \exp(i\dfrac{\omega n L}{2 c})}{1-R \:\exp(i \dfrac{2\omega n L}{c} )}
	\end{split}
\end{equation}
where $R$ is the modal reflectivity (in intensity) of the SPDC modes, $n$ their modal refractive index and $L$ is the waveguide length. We assume for simplicity that $R$ and $n$ are the same for both polarizations (which is correct within 5 \% for our sample).
Each waveguide facet is then modeled as a frequency-dependent beamsplitter \cite{Ou07}, where photons can either be reflected or transmitted  with a probability depending on their frequency. The corresponding transformations for the $\hat{a}$ and $\hat{a}^{\dagger}$ operators read:
\begin{equation}
	\begin{split}
		& \hat{a}^{\dagger}_{s, \sigma}(\omega) \rightarrow f_t(\omega)\hat{a}^{\dagger}_{\mathcal{R},  \sigma}(\omega) + f_r(\omega)\hat{a}^{\dagger}_{\mathcal{L},  \sigma}(\omega)\\
		& \hat{a}^{\dagger}_{i,  \sigma}(\omega) \rightarrow f_t(\omega)\hat{a}^{\dagger}_{\mathcal{L}, \sigma}(\omega) + f_r(\omega)\hat{a}^{\dagger}_{\mathcal{R},  \sigma}(\omega)
	\end{split}
	\label{eq:BS}
\end{equation}
where $ \sigma$ stands for the $H$ or $V$ polarization. Starting from Eq. \eqref{Eq:Psi}, applying transformations \eqref{eq:BS} and adding the effect of the delay line (delay $\tau$) leads to the following expression for the biphoton state just before the beamsplitter:
\begin{widetext}
	\begin{equation}
	\begin{split}
		\ket{\Psi} =  \iint \dd\omega_1\dd\omega_2 \biggl[
		\phi_{HV} (\omega_1, \omega_2) & \bigl(f_t(\omega_1) \hat{a}^{\dagger}_{\mathcal{R},H}(\omega_1) e^{-i\omega_1\tau} + f_r(\omega_1)\hat{a}^{\dagger}_{\mathcal{L},H}(\omega_1) \bigr)  \\
		\cdot & \bigl(f_t(\omega_2) \hat{a}^{\dagger}_{\mathcal{L},V}(\omega_2) + f_r(\omega_2)\hat{a}^{\dagger}_{\mathcal{R},V}(\omega_2) e^{-i\omega_2\tau} \bigr) \\
		+ \: \phi_{VH} (\omega_1, \omega_2) & \bigl(f_t(\omega_1) \hat{a}^{\dagger}_{\mathcal{R},V}(\omega_1) e^{-i\omega_1\tau} + f_r(\omega_1)\hat{a}^{\dagger}_{\mathcal{L},V}(\omega_1) \bigr) \\
		\cdot & \bigl(f_t(\omega_2) \hat{a}^{\dagger}_{\mathcal{L},H}(\omega_2) + f_r(\omega_2)\hat{a}^{\dagger}_{\mathcal{R},H}(\omega_2) e^{-i\omega_2\tau} \bigr) \biggr] \ket{0,0}
	\end{split}
\end{equation}
\end{widetext}
We then apply the usual beamsplitter transformations:
\begin{equation}
	\begin{split}
		\hat{a}^\dagger _1(\omega) \rightarrow \dfrac{1}{\sqrt{2}} \qty( \hat{a}^\dagger _3(\omega) + i \hat{a}^\dagger _4(\omega) ) \\
		\hat{a}^\dagger _2(\omega) \rightarrow \dfrac{1}{\sqrt{2}} \qty( \hat{a}^\dagger _4(\omega) + i \hat{a}^\dagger _3(\omega) )
	\end{split}
	\label{Eq:Cat, BsTransformation}
\end{equation}
and since we are interested in coincidence events, only the crossed terms (of the kind $\hat{a}^\dagger _3\hat{a}^\dagger _4$ and $\hat{a}^\dagger _4\hat{a}^\dagger _3$) are considered. The resulting effective wavefunction $\ket{\Psi_\text{c}}$ reads:
\begin{equation}
	\begin{split}
		\ket{\Psi_\text{c}}=\dfrac{1}{2} \iint \dd\omega_1\dd\omega_2 \: \bigl[ & A(\omega_1, \omega_2) \hat{a}^\dagger_{3, H}(\omega_1)\hat{a}^\dagger_{4, V}(\omega_2)   \\
		+ & B(\omega_1, \omega_2) \hat{a}^\dagger_{3, H}(\omega_2)\hat{a}^\dagger_{4, V}(\omega_1)  \\
		+ & C (\omega_1, \omega_2) \hat{a}^\dagger_{3, V}(\omega_1)\hat{a}^\dagger_{4, H}(\omega_2)  \\
		+ & D(\omega_1, \omega_2) \hat{a}^\dagger_{3, V}(\omega_2)\hat{a}^\dagger_{4, H}(\omega_1) \bigr] \ket{0,0}
			\end{split}
\end{equation}
where the coefficients are :
\begin{align}
	\begin{split}
		A(\omega_1, \omega_2)= & \phi_{HV}(\omega_1, \omega_2)
		  \bigl[- f_r(\omega_1)f_r(\omega_2) e^{-i\omega_2\tau} \\
		 & + i f_t(\omega_1) f_r(\omega_2) e^{-i(\omega_1+\omega_2)\tau} \\
		 & + i f_r(\omega_1) f_t(\omega_2) + i f_t(\omega_1) f_t(\omega_2) e^{-i\omega_1\tau}\bigr]
	\end{split}
	\\	
	\begin{split}
		B(\omega_1, \omega_2)= & \phi_{VH}(\omega_1, \omega_2) 
		  \bigl[f_r(\omega_1)f_r(\omega_2) e^{-i\omega_2\tau} \\
		  & + i f_t(\omega_1) f_r(\omega_2) e^{-i(\omega_1+\omega_2)\tau} \\
		& + i f_r(\omega_1) f_t(\omega_2) - i f_t(\omega_1) f_t(\omega_2) e^{-i\omega_1\tau}\bigr]
	\end{split}
	\\
	\begin{split}
		C(\omega_1, \omega_2)= & \phi_{VH}(\omega_1, \omega_2)
		 \bigl[- f_r(\omega_1)f_r(\omega_2) e^{-i\omega_2\tau} \\
		 & + i f_t(\omega_1) f_r(\omega_2) e^{-i(\omega_1+\omega_2)\tau} \\
		& + i f_r(\omega_1) f_t(\omega_2) + i f_t(\omega_1) f_t(\omega_2) e^{-i\omega_1\tau}\bigr]
	\end{split}
	\\
	\begin{split}
		D(\omega_1, \omega_2)= & \phi_{HV}(\omega_1, \omega_2) 
		 \bigl[f_r(\omega_1)f_r(\omega_2) e^{-i\omega_2\tau} \\
		 & + i f_t(\omega_1) f_r(\omega_2) e^{-i(\omega_1+\omega_2)\tau} \\
		& + i f_r(\omega_1) f_t(\omega_2) - i f_t(\omega_1) f_t(\omega_2) e^{-i\omega_1\tau}\bigr]
	\end{split}
\end{align}

In general, the coincidence probability $P_c$ would consist of four contributions corresponding to the detection of $HH$, $VV$, $HV$, or $VH$ photons at the beasmplitter output. However, even if the photons are mixed by the cavity, they are always generated with crossed polarizations so that the $HH$ and $VV$ events can be neglected. The coincidence probability for the $HV$ and $VH$ events can be calculated as the expectation value of the coincidences operators $\hat{M}_{HV}$ and $\hat{M}_{VH}$:
\begin{equation}
	\begin{split}
		 \hat{M}_{HV}= & \int \dd\omega_3 \hat{a}^{\dagger}_{3,H}(\omega_3)\ket{0}\bra{0}\hat{a}_{3,H}(\omega_3) \\
		& \cdot \int \dd\omega_4 \hat{a}^{\dagger}_{4,V}(\omega_4)\ket{0}\bra{0}\hat{a}_{4,V}(\omega_4)
		\\
		 \hat{M}_{VH}= & \int \dd\omega_3 \hat{a}^{\dagger}_{3,V}(\omega_3)\ket{0}\bra{0}\hat{a}_{3,V}(\omega_3) \\
		& \cdot \int \dd\omega_4 \hat{a}^{\dagger}_{4,H}(\omega_4)\ket{0}\bra{0}\hat{a}_{4,H}(\omega_4)
	\end{split}
\end{equation}
The resulting coincidence probabilities read:
\begin{align}
	\begin{split}
		P_{HV}(\tau) =& \ev{\hat{M}_{HV}}{\Psi_\text{c}} \\
		=& \dfrac{1}{4}\iint \dd \omega_3 \, \dd\omega_4 \bigl[  \vert A(\omega_3, \omega_4)\vert^2 \\
		& +  \vert B(\omega_3, \omega_4)\vert^2 + A^*(\omega_3, \omega_4) B(\omega_4, \omega_3) \\
		 &+  A(\omega_4, \omega_3) B^*(\omega_3, \omega_4) \bigr]
	\end{split}
	\\
	\begin{split}
		P_{VH}(\tau) =& \ev{\hat{M}_{VH}}{\Psi_\text{c}}\\
		  =& \dfrac{1}{4}\iint  \dd \omega_3 \, \dd\omega_4 \bigl[ \vert C(\omega_3, \omega_4)\vert^2 \\ 
		&+ \vert D(\omega_3, \omega_4)\vert^2 +  C^*(\omega_3, \omega_4) D(\omega_4, \omega_3) \\
		& + C(\omega_4, \omega_3) D^*(\omega_3, \omega_4) \bigr]
	\end{split}
\end{align}
which leads to the total coincidence probability $P_c(\tau)=P_{HV}(\tau)+P_{VH}(\tau)$.

\begin{figure*}[t]
	\centering
	\includegraphics[width=0.65\textwidth]{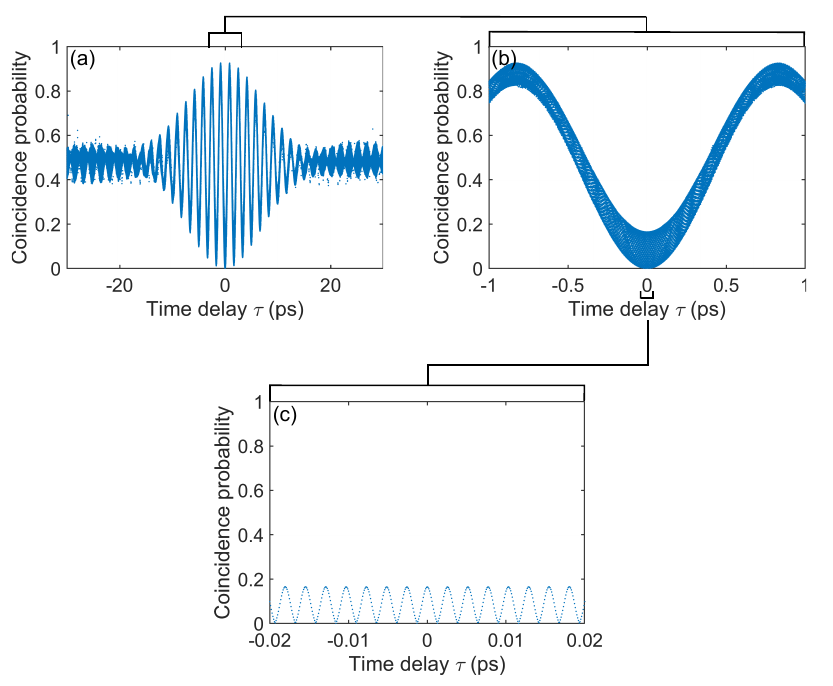}
	\caption{Simulated HOM interferogram for the HPF entangled state, taking into account the Fabry-Perot effect of the sample with facet reflectivity $R=10\%$. From (a) to (c) the time axis is zoomed around $\tau=0$, in order to show the additional modulation at the pump frequency. The scattered points are caused by numerical artifacts in the integration.}
	\label{HOM_Reflectivity}
\end{figure*}

We numerically evaluate $P_c(\tau)$ by considering Gaussian phase-matching functions (Eq. \eqref{Eq:PM}) with parameters given by the experiment and facet reflectivity $R=0.1$. Figure \ref{HOM_Reflectivity} shows the calculated HOM interferogram for different levels of close-up around $\tau=0$. The general shape of the coincidence probability (Fig. \ref{HOM_Reflectivity}a) is similar to the case without cavity effect (Fig. \ref{Fig2}b), with a Gaussian-like envelope and a sinusoidal oscillation. However, zooming in we note another oscillation superimposed to the first one (Fig. \ref{HOM_Reflectivity}b and c). This modulation has a period of 2.3 fs, corresponding to the inverse of the pump frequency ($2\pi / \omega_p$).

\begin{figure}
	\centering
	\includegraphics[width=0.65\columnwidth]{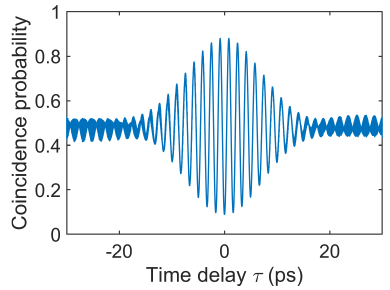}
	\caption{Simulated HOM interfogram obtained from the data of Fig. \ref{HOM_Reflectivity} by averaging out the modulation at the pump frequency, leading to a decrease of effective fringe visibility.}
	\label{HOM_Reflectivity_Average}
\end{figure}

As stated in the main text, the used HOM experimental setup does not have enough resolution (i.e. mechanical and thermal stability of the mirrors and of the delay line) to resolve this oscillation at the pump frequency, and thus only its temporal average is measured. The latter is numerically evaluated in Fig. \ref{HOM_Reflectivity_Average}. We observe a reduction of the effective visibility of the fringes to $\simeq 82 \%$.
As mentioned in the conclusion of the article, using a multi-layer instead of single-layer coating could decrease the facet reflectivities to $\leqslant 1 \%$, for which the theoretical HOM visibility would be $\geqslant 98 \% $ (all other conditions assumed to be perfect).

\section*{Acknowledgements}

We acknowledge support from European Union’s Horizon 2020 research and innovation programme under the Marie Skłodowska-Curie grant agreement No 665850, Paris Ile-de-France Région in the framework of DIM SIRTEQ (LION project), Ville de Paris Emergence program (LATTICE project), Labex SEAM (Science and Engineering for Advanced Materials and devices, ANR-10-LABX-0096), IdEx Université de Paris (ANR-18-IDEX-0001), and the French RENATECH network.

\section*{Disclosures}

 The authors declare no conflicts of interest.

\bibliography{D:/Travail/MPQ/Biblio}

\end{document}